\documentclass[9pt,twocolumn,twoside]{osajnl}

\journal{josab} 

\setboolean{shortarticle}{false} 

\usepackage{subfig}
\title{Recent Advances in Supercontinuum  Generation in Specialty Optical Fibers}
\author[1,*]{T. Sylvestre}
\author[1,2]{E. Genier}
\author[1,3]{A. N. Ghosh}
\author[4]{P. Bowen}
\author[5]{G. Genty}
\author[6]{J. Troles}
\author[2]{A. Mussot}
\author[3]{A. C. Peacock}
\author[7]{M. Klimczak}
\author[8]{A. M. Heidt}
\author[9]{J. C. Travers}
\author[4,10]{O. Bang}
\author[1]{J. M. Dudley}

\affil[1]{Institut FEMTO-ST, UMR 6174 CNRS-Universit\'{e} Bourgogne Franche-Comt\'{e}, 25030 Besan\c{c}on, France}
\affil[2]{Univ. Lille, CNRS, UMR 8523-PhLAM-Physique des Lasers Atomes et Molécules, F-59000 Lille, France}
\affil[3]{Optoelectronics Research Centre, University of Southampton, Southampton, SO17 1BJ, UK}
\affil[4]{NKT Photonics A/S, Blokken 84, DK-3460, Birker\o d, Denmark}
\affil[5]{Photonics Laboratory, Tampere University, FI-33104 Tampere, Finland}
\affil[6]{Université de Rennes, CNRS, ISCR-UMR 6226, 35000 Rennes, France}
\affil[7]{Faculty of Physics, University of Warsaw, Pasteura 5, 02-093 Warsaw, Poland}
\affil[8]{Institute of Applied Physics, University of Bern, Switzerland}
\affil[9]{School of Engineering and Physical Sciences, Heriot-Watt University, Edinburgh, EH14 4AS, UK}
\affil[10]{DTU Fotonik, Department of Photonics Engineering, Technical University of Denmark, 2800 Kgs. Lyngby, Denmark}
\affil[*]{Corresponding author: thibaut.sylvestre@univ-fcomte.fr}

\begin{abstract}
The physics and applications of fiber-based supercontinuum (SC) sources have been a subject of intense interest over the last decade, with significant impact on both basic science and industry. New uses for SC sources are also constantly emerging due to their unique properties that combine high brightness, multi-octave frequency bandwidth, fiber delivery and single-mode output. The last few years have seen significant research efforts focused on extending the wavelength coverage of SC sources towards the 2 to 20~$\mu$m molecular fingerprint mid-infrared (MIR) region and in the ultraviolet (UV) down to 100~nm, while also improving stability, noise and coherence, output power and polarization properties. Here we review a selection of recent advances in SC generation in a range of specialty optical fibers including: fluoride, chalcogenide, telluride, and silicon-core fibers for the MIR; UV-grade silica fibers and gas-filled hollow-core fibers for the UV range; and all-normal dispersion fibers for ultra-low noise coherent SC generation.\\

\end{abstract}

\setboolean{displaycopyright}{true}

\begin{document}

\maketitle

\section{Introduction}
\label{Intro}
A longstanding challenge since the invention of the laser has been the development of efficient methods to 
exploit nonlinearity to convert laser light to new wavelengths. Supercontinuum (SC) generation in optical fibers has been shown to offer a convenient and elegant solution to this challenge, as it massively broadens the laser spectrum while maintaining a spatially coherent output \cite{Alfano_Book_2016,Dudley_book_2010}. This provides an inherently fiber-delivered broadband light spectrum possessing the brightness of a laser and the spectral width of a lamp, capable of replacing most light sources used today in optical metrology, spectroscopy, and microscopy. Important applications of fiber-based broadband SC sources include bio-imaging, optical coherence tomography (OCT), material processing, optical sensing, absorption spectroscopy, and optical frequency comb (OFC) technologies \cite{Alfano_Book_2016,Dudley_book_2010,Genty_JOSAB_2007,Udem_Nat_2002,Labruyere_OFT_2012}. Many applications also benefit from the fact that SC light is several orders of magnitude spectrally brighter than blackbody radiation and its coherence properties  enable the use of relatively simple interferometric and heterodyne detection schemes.

\begin{figure*}[ht!]
\centering
\includegraphics[width=18cm]{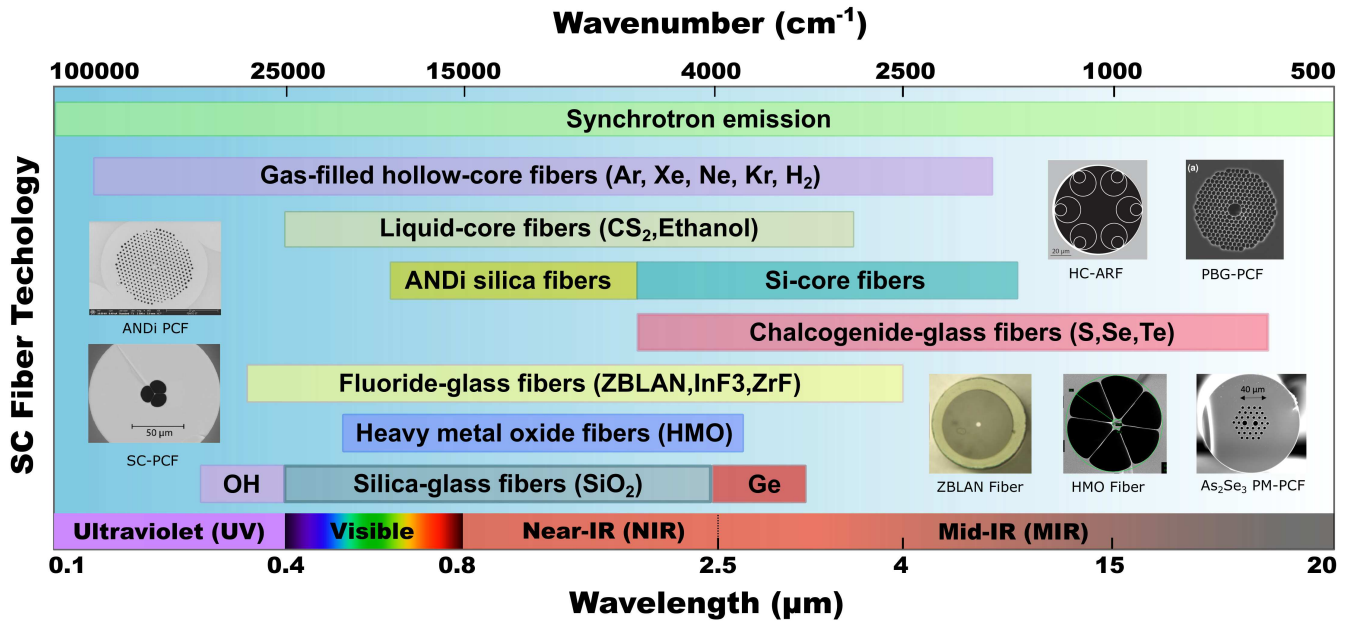}
\caption{A survey of SC bandwidths in various specialty optical fibers from the ultraviolet (UV) to the MIR ranges. The insets show some examples of the different fiber platforms used for SC generation including (left) all-normal dispersion (ANDi) PCF and suspended-core PCF, (top right) hollow-core  anti-resonant fiber (HC-ARF) and photonic band gap (PBG-PCF), (bottom right) fluoride-glass (ZBLAN) step-index fiber, Heavy-metal-oxyde (HMO-PCF) and chalcogenide (AS$_2$Se$_3$) polarization-maintaining (PM) PCF.}
\label{SCreview}
\end{figure*}
State-of-the-art commercial SC fiber sources are mostly based on silica-glass photonic crystal fibers (PCFs), providing watt-level output power over the complete silica fiber transmission window, i.e. from 400~nm to 2400~nm. Fluoride-fiber based mid-infrared (MIR) SC sources with spectra reaching up to 4~$\mu$m (in some cases up to 4.8~$\mu$m) have also become commercially available \cite{LEUKOS_2021,NKT_2021}.  However, many emerging applications require extension of the SC spectrum beyond the state-of-the-art, towards both the UV range (ideally from 50 to 400~nm) and the MIR range from 2 to 20~$\mu$m.  There is particular need for broadband sources of UV light for fluorescence imaging and for diagnostic and therapeutic uses in medicine \cite{Poudel-JOSAB-2019}, and the MIR covers the molecular fingerprint region and the atmospheric windows of 3–5 $\mu$m and 8–12 $\mu$m \cite{Petersen_NP_2014,Wang-PX-2021}. Covering this MIR range is necessary for OCT imaging, remote sensing, air pollution monitoring, homeland security, and minimally invasive medical surgery \cite{Hancock_JPC_2008,Manninen_OE_2014,Israelsen_LSA_2019,Zorin_OE_2018,Sych_OE_2010,Hult_OE_2007,Amiot_APL_2017,Troles-2020,Lemiere-2021}.

 Recent years have seen significant progress in overcoming the current shortcomings of SC sources in terms of wavelength coverage, noise, power density and robustness.  Significant developments have been made in reaching target UV and MIR wavelength ranges, and the fiber SC has matured considerably to become a truly disruptive technology able to meet a range of societal and industrial challenges.

A key reason for these advances has been the development of new classes of specialty optical fiber, and the purpose of this paper is to present a review of this work.  The paper is organized as follows. The second section introduces the physics of SC generation and presents an overview of the approaches and  techniques used for  fiber-based SC generation. In Section \ref{MIR}, we provide a brief review of recent results in MIR SC generation in soft-glass fibers using direct and cascaded pumping. We specifically describe work developing compact and reliable MIR SC sources based on fiber laser pumped cascaded fiber systems. A short segment is then devoted to the new silicon-core fibers for SC applications. Whilst IR fiber SC technology is becoming more and more robust, SC generation in the UV is still a developing area. In Section \ref{UV}, we describe the different specialty fibers for UV SC generation, including gas-filled hollow-core fibers and UV-grade silica fibers. In Section \ref{ANDI}, we discuss  recent achievements developing low-noise and coherent SC sources based on ANDi PCFs, as well as presenting an overview of  potential applications. 

To introduce the topics covered in this review, it is useful to begin by presenting a summary of the various SC bandwidths that have been reported using different optical fiber platforms. Figure \ref{SCreview} presents a survey of fiber SC spectral coverage from the UV (100~nm) to the MIR ranges up to 20~$\mu$m. These include: standard silica-based PCF SC source from 400~nm to 2.4~$\mu$m, UV-grade (High-OH-dopant) silica fibers (OH) for UV extension down to 300~nm \cite{Perret_OSA_2020}, highly GeO$_2$-doped silica fibers (Ge) for IR extension up to 3.2~$\mu$m \cite{Jain-OPEX-2016}, respectively. The survey also shows the SC bandwidths achieved using heavy-metal oxide fibers (HMO), fluoride-glass fibers up to 4.8~$\mu$m, chalcogenide-glass fibers up to 18~$\mu$m, liquid-core and gas-filled hollow-core fibers. The insets show scanning electron microscope (SEM) images of some of the different PCF platforms used for SC generation including: (left) ANDi PCF and suspended-core PCF, (top right) HC-ARF and PBG-PCF, (bottom right) ZBLAN fiber, HMO-PCF and chalcogenide (AS$_2$Se$_3$) PCF.

\section{Supercontinuum Physics}
\label{Physics}
There have been many attempts to control and harness nonlinear and spectral broadening dynamics in silica fibers \cite{Agrawal}, but it was truly the advent of the microstructured air-hole fiber (PCF) in the mid 90s that revolutionized the generation of a broadband SC \cite{Russel_JLT_2006,Ranka_OL_2000}. In particular, PCFs have allowed tailoring of the fiber dispersion while increasing the efficiency of nonlinear effects leading to SC spectra extending over the full transmission window of silica fibers, and enabling applications outside laboratories. A key aspect in the success of PCFs for SC generation was to match the zero-dispersion wavelength with the wavelength of available ultrafast laser systems, enabling efficient seeding of soliton dynamics that underpin the spectral broadening \cite{Dudley_RMP_2006}. 

\begin{figure}[h!]
\centering
\includegraphics[width=8.8 cm]{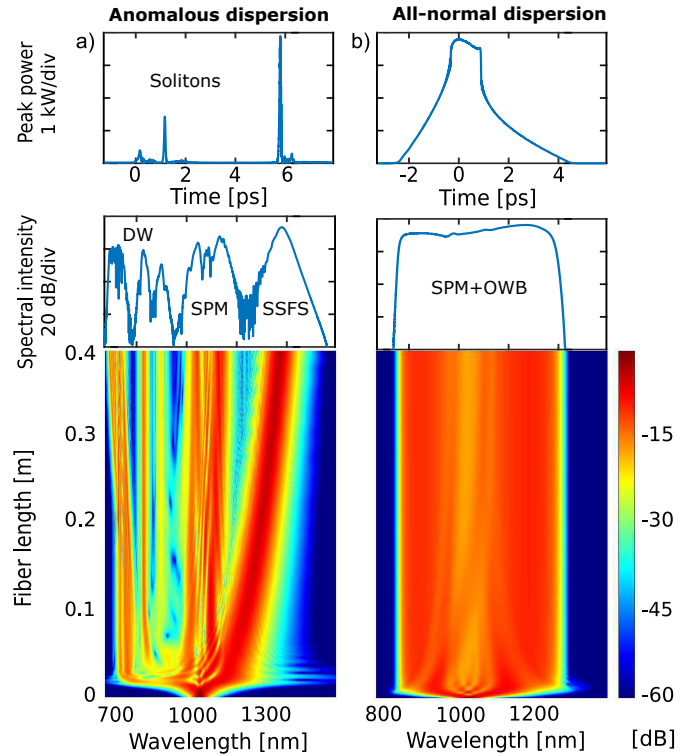}
\caption{Numerical simulations based on the generalized nonlinear Schrödinger equation of (a) anomalous and (b) ANDi SC generation as a function of propagation distance in the fiber. The input parameters are: 50 fs input pulse duration at 1040 nm with 10 kW input peak power and $\beta_2$ = -11 ps$^2.km^{-1}$ for (a) and with 50 kW input peak power and $\beta_2$ = 5.27 ps$^2.km^{-1}$ for (b).}
\label{ANDiSC}
\end{figure}

Temporal solitons are remarkably robust nonlinear optical waves formed under the proper balance between linear dispersion and nonlinear self-phase modulation (SPM). Exciting soliton dynamics requires injecting short pulses in the anomalous dispersion regime of a fiber, which in most cases corresponds to pumping at wavelengths longer than the fiber zero-dispersion wavelength (ZDW). In this regime, the input pulse experiences an initial stage of higher-order soliton compression and fission into fundamental solitons. Perturbation of the solitons dynamics by higher-order dispersion and stimulated Raman scattering (SRS) leads to the emission of dispersive waves (DW) in the visible and soliton self-frequency shift (SSFS) towards the infrared \cite{Dudley_RMP_2006}. A numerical illustration of these processes is shown in Fig. \ref{ANDiSC}(a) which plots the typical spectral evolution of a 50~fs input hyperbolic secant pulse at 1040~nm in the anomalous dispersion regime of a single-mode silica PCF. The simulation is based on the general nonlinear Schrödinger equation (GNLSE) solved with the split-step method \cite{Dudley_RMP_2006} and using parameters as given in the figure caption.
The nonlinear dynamics are primarily governed by the fiber dispersion profile and a wide variety of dispersion-engineered PCFs have been designed and fabricated for to match the ZDW with various pump lasers around 800 nm, 1040 nm, 1064 nm, 1550 nm, 2 µm and beyond. Much effort has also been dedicated to design PCFs with two ZDWs (parabolic dispersion), enabling solitons to emit far-detuned DWs on both sides of the SC spectrum \cite{Mussot_OE_2007}.

Although soliton-based SC generation is known to yield the broadest SC spectra, SC generation in ANDi fibers developing from SPM and optical wave breaking (OWB) has also been extensively studied. ANDi SC requires higher peak power femtosecond pulses and leads to pulse-preserved, flat-top, fully coherent SC spectra as confirmed by numerical simulations (see Fig. \ref{ANDiSC}(b)) \cite{Heidt2016}.
These features make ANDi SC of particular interest for applications requiring high coherence as well as uniform and smooth spectral and temporal intensity profiles. However, despite relative abundance of results on ANDi PCF and coherent SC generation demonstrations in the literature (see Section 5 for a detailed review), there is a lack of detailed experimental characterization of the spectro-temporal properties of ANDi SC pulses. A recently published study exploiting time domain-ptychography for full-field characterization \cite{Heidt2016b}, has revealed the importance of using high quality pump pulses for the generation of stable SC. In particular, this study has shown that ANDi SC pulses may exhibit complex temporal fine structure on time scales of a few optical cycles arising from the nonlinear amplification of imperfections of the pump pulse such as pre-/post-pulses and low-level pedestals \cite{Rampur2020}. These results somewhat contrast the general perception that ANDi SC inherently possess continuous and uniform spectrograms as typically seen in numerical simulations assuming ideal femtosecond pump pulses \cite{Heidt2010}. However, taking into account the full-field characteristics of the pump pulses (e.g. retrieved from a frequency-resolved optical gating measurement) in GNLSE simulations enabled the reconstruction of the observed ANDi SC pulse complexity in remarkable agreement with the experimental measurements. It further enabled linking the observed temporal fine structure features to specific fiber parameters such as birefringence \cite{Rampur2020}.

SC generation in silica PCFs has been observed using not only ultra-short femtosecond seed pulses but also using longer picosecond or nanosecond pulses \cite{Coen_OL_2001,Dudley_JOSAB_2002,Wadsworth_OE_2004}, and even in the continuous-wave (CW) regime \cite{Abeeluck_OL_2004,Sylvestre_OL_2006,Travers_OE_2008,Kudlinski_OL_2008}. In the long pulse regime, the key nonlinear process underlying SC generation in the anomalous dispersion region is noise-seeded modulation instability (MI) which breaks up the input pulse envelope into a train of hundreds sub-picosecond breather pulses with random characteristics (amplitude and duration) that subsequently undergo soliton dynamics similar to those observed in the short pulse regime \cite{Dudley_OE_2009}. In the normal dispersion regime, the SC develops from the combined effects of noise-seeded parametric four-wave mixing (FWM) processes assisted by cascaded Raman scattering (CRS) \cite{Mussot_OL_2003,Lesvigne_OL_2007,Perret_APL_2019}. In multimode or birefringent fibers, the FWM processes can originate from phase-matched intermodal and vectorial nonlinear instabilities including polarization MI (PMI) and cross-phase MI (XPMI), respectively. Dual-pumping schemes have been also implemented for multi-octave spanning SC generation based on cross-phase modulation (XPM) \cite{Champert_OE_2004}.

More recently, SC generation has also been reported in graded-index (GRIN) multimode optical fibers, triggered by geometric parametric instability due to periodic self-imaging arising from multimodal interference, and which can under particular injection conditions lead to the generation of broadband self-cleaned beam output \cite{Lopez-Galmiche_OL_2016,Krupa_PRL_2017,Dupiol_OL_2017}. 
SC generation in liquid-core optical fibers (LiCOF) has also seen increased interest in recent years due to the high refractive index and wide MIR spectral transparency of inorganic solvents. A particular feature associated with liquids from the nonlinear physics viewpoint is their non-instantaneous nonlinearity and stimulated nonlinear scattering arising from the slow molecular motion of liquids, characteristics which gives rise to specific nonlinear dynamics and allows for highly-coherent SC generation \cite{Chemnitz_NC_2017,Chemnitz_Optica_2018,Juniad_OE_2021,Fanjoux_JOSAB_2017}.

\section{MIR supercontinuum generation}
\label{MIR}

SC generation in the MIR regime has been extensively investigated in the past 15 years with more than 400 scientific papers published on this topic. It is still a very active field of research, driven by a wide range of potential applications including molecular spectroscopy, OCT, material processing, and optical sensing \cite{Petersen_NP_2014,Wang-PX-2021,Hancock_JPC_2008,Manninen_OE_2014,Israelsen_LSA_2019,Zorin_OE_2018,Sych_OE_2010,Hult_OE_2007,Amiot_APL_2017,Troles-2020,Lemiere-2021}. 

\subsection{Soft-glass fibers}

The development of MIR SC sources poses a number of important challenge due to requirement for non-silica soft glasses, complex fabrication techniques, and new pump laser wavelengths. Significant efforts have been devoted to the synthesis of soft glasses for the MIR \cite{Tao_AOP_2015,Meneghetti_2019}, including chalcogenide arsenic trisulfide (As$_2$S$_3$), arsenic triselenide (As$_2$Se$_3$), germanium arsenic selenide (GeAsSe), germanium telluride (GeTe and GeAsTeSe) \cite{Petersen_NP_2014,Petersen_OE_2017,LUCAS_OMER_2013}, tellurite (TeO$_2$) \cite{Des_2017,Saini_2020}, chalcohalides (Ge–Te–AgI) \cite{,Zhao_LPR_2017}, heavy metal oxide (PbO-Bi$_2$O$_3$-Ga$_2$O$_3$-SiO$_2$-CdO) \cite{Ghosh_JOSAB_2018} and fluorides (ZBLAN:ZrF$_4$ –BaF$_2$ –LaF$_3$ –AlF$_3$ –NaF; InF$_3$; ZrF$_4$) \cite{Xia_OL_2006,Agger_JOSAB_2012,Michalska_OJ_2017}. Among the large variety of infrared fibers, chalcogenide-glass-based fibers (composed of chalcogen elements such as S, Se or Te) are excellent platforms for SC applications in the MIR due to their wider transmission window, tailorable dispersion, and hundred times larger nonlinearity compared to silica or ZBLAN fibers \cite{Petersen_NP_2014,Cheng_OL_2016,Petersen_OE_2017,Theberge_OE_2018,Lemiere_JOSAB_2019}. Bulk chalcogenide glasses are usually prepared using several techniques such as melt-quenching \cite{Shiryaev_OM_2015,Meneghetti_2019} or microwave radiation \cite{Sivakumaran_JPAP_2005,Thompson_PCG_2013} that can be drawn into highly nonlinear step-index fibers \cite{Shiryaev_OM_2015} or dispersion-tailored PCFs using techniques such as molding  \cite{Coulombier_2010,Caillaud_OE_2016,Ghosh_JPP_2019}, drilling \cite{Amraoui_OE_2010} and extrusion \cite{Jiang_2017}. 

Below, we present a brief review of the state-of-the-art for MIR SC generation in soft-glass fibers, classified into direct pumping and cascaded pumping. Direct pumping is based on bulky and expensive mid-IR pump laser sources such as tunable optical parametric oscillators (OPOs) and optical parametric amplifiers (OPAs), which can suffer from low output power and poor stability. Cascaded systems based on fiber lasers and MIR nonlinear fibers have recently emerged as attractive and promising alternatives, opening routes towards practical, table-top, and robust MIR SC sources with high spectral power density. Of particular interest are compact and reliable all-fiber systems pumped with standard fiber lasers at telecommunication wavelengths where an initial pulse at a wavelength of 1550 nm is gradually shifted towards longer wavelengths in a cascade of silica and soft-glass fibers, enabling a step wise extension to the MIR. From a fundamental physics viewpoint, the cascade allows for strongly enhancing the SSFS dynamics using dispersion-engineered and highly nonlinear fiber segments, enabling extension of the SC far in the MIR \cite{Kubat_OE_2014}.

\begin{figure}[ht!]
\centering
\includegraphics[width=8.8 cm]{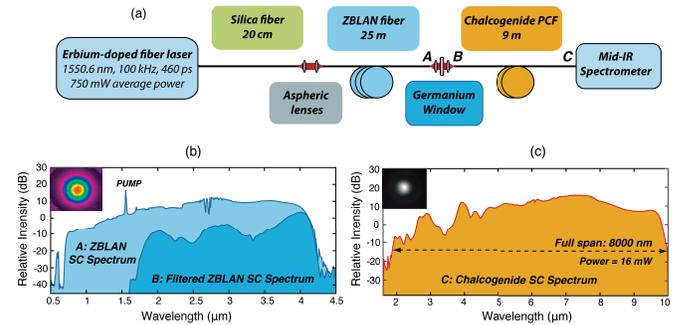}
\caption{a) Experimental setup for mid-infrared SC generation in a cascaded silica-ZBLAN-chalcogenide optical fiber system. b) Experimental SC spectra
at the ZBLAN fiber output (light blue) and after the long-pass Germanium filter (dark blue). c) Experimental SC spectrum at the chalcogenide fiber output (yellow). The insets show the optical mode profiles out of both the ZBLAN (b) and chalcogenide (c) optical fibers. Taken from Ref. \cite{Venck_LPR_2020}.}.
\label{CascadeSC}
\end{figure}
Among the direct pumping SC generation systems, Cheng et al. have reported the generation of a SC spanning from 2-15.1~$\mu$m in a 3~cm long step-index As$_2$Se$_3$ fiber \cite{Cheng_OL_2016}. The fiber with a core diameter of 15~$\mu$m was pumped at 9.8 $\mu$m with a peak power of 2.89~MW delivered from a MIR pump source comprising of a Ti:Sapphire mode-locked seed laser, a traveling-wave optical parametric amplifier of superfluorescence (TOPAS), and a difference frequency generation (DFG) unit \cite{Cheng_OL_2016}. Ou et al. also reported a MIR SC spectrum spanning from 1.8 to 14~$\mu$m in a 20~cm long and 23~$\mu$m core diameter Ge$_{15}$Sb$_{25}$Se$_{60}$/Ge$_{15}$Sb$_{20}$Se$_{65}$ step-index fiber pumped by 150 fs pulses at 6~$\mu$m delivered by an OPA system \cite{OU_OL_2016}. Using an ultra-high numerical aperture chalcogenide step-index fiber, C. R. Petersen et al. demonstrated a SC spectrum covering from 1.4 to 13.3~$\mu$m, and with an output power of 150 $\mu$W with potential for upscaling by increasing the repetition rate or the pulse duration \cite{Petersen_NP_2014}. The spectral broadening was achieved by pumping a 16 $\mu$m core diameter fiber made from As$_{40}$Se$_{60}$/Ge$_{10}$As$_{23.4}$Se$_{66.6}$ (core/cladding) glass at 6.3 $\mu$m using a Ti:sapphire laser pumped tunable OPA and a noncollinear DFG module. Yu et al. reported the widest spectrum in step-index chalcogenide fibers with a low peak power of 3 kW \cite{Yu_OL_2015}. By pumping an 11 cm long Ge$_{12}$As$_{24}$Se$_{64}$/Ge$_{10}$As$_{24}$S$_{66}$ fiber with 330 fs pulses at 4.0 $\mu$m emitted from an OPA system, they obtained a SC spectrum spanning from 1.8 to 10 $\mu$m \cite{Yu_OL_2015}.

Many chalcogenide fibers contain elements such as arsenic and antimony which can be associated with possible safety issues during glass synthesis, fiber drawing, and testing. Research has thus also been devoted to remove arsenic and antimony elements from chalcogenide glasses, while also extending the glass transmission window to beyond 20 $\mu$m. Lemière et al. recently demonstrated a MIR SC in arsenic and antimony free chalcogenide glass step-index fiber \cite{Lemiere_RiP_2021}. Specifically, they obtained 1.7-18~$\mu$m SC generation in a  40 mm long and 12 $\mu$m core Ge-Se-Te fiber pumped at 8.15 $\mu$m with 65 fs pulses delivered by a  non-collinear OPA followed by a DFG module \cite{Lemiere_RiP_2021}. Zhao et al. reported a MIR SC spanning from 2 to 16 $\mu$m in a 14 cm long low-loss step-index telluride (Ge–Te–AgI) fiber pumped in the normal dispersion regime at 7 $\mu$m by 150 fs pulses emitted from an OPA system \cite{Zhao_LPR_2017}. Lemière et al. also reported SC generation spanning from 2 to 14 $\mu$m in a 40 mm long 10 $\mu$m core Ge$_{20}$Se$_{60}$Te$_{20}$/Ge$_{20}$Se$_{70}$Te$_{10}$ step-index fiber \cite{Lemiere_JOSAB_2019}. Significant progress among the direct pumping schemes has been achieved in 2017 by Hudson et al. using as a pump source a holmium-based fiber laser at 2.9 $\mu$m \cite{Hudson_Optica_2017}. They demonstrated 1.8-9.5 $\mu$m MIR SC with an average output power of around 30 mW in a polymer-coated chalcogenide fiber taper with this MIR fiber laser. More recently, a record power of 825 mW has been reached using an As$_2$S$_3$ fiber-based SC source between 2.5 and 5.0 $\mu$m, with Al$_2$O$_3$ anti-reflection coatings that were sputtered on chalcogenide fiber tips to increase the launching efficiency up to 82\%, making this record output power possible \cite{Robichaud_2020}.

\begin{table*}[ht!]
\caption{\label{table1} List of recently developed cascaded and all-fiber systems for MIR SC generation.}
\begin{tabular}{ccccccc}
Ref. & Fiber systems & Amplifiers & SC bandwidth & Power
 & Rep. rate & pulse duration  \\ \hline
\cite{Xia_OL_2006} & silica-ZBLAN fibers
& No & 0.8 - 4.5 $\mu$m & 23 mW & 5 kHz & 2 ns \\
\cite{Venck_LPR_2020} & Silica-ZBLAN-As$_2$Se$_3$ fibers
& No & 2 - 10 $\mu$m & 16 mW & 100 kHz & 460 ps \\
\cite{Hudson_Optica_2017} & Chalcogenide taper & No &
1.8 - 9.5 $\mu$m & 30 mW & 42 MHz & 230 fs \\
\cite{Xia_OL_2007} & Silica-ZBLAN fibers & Yes &1 - 3.2 $\mu$m & 1.3 W & 5 kHz & 2 ns \\
\cite{Hagen_PTL_2006} & Silica-ZBLAN fibers &No &1.8 - 3.4 $\mu$m &5 mW
&200 kHz & 900 fs \\
\cite{Woyessa_OL_2021} &  Doped Silica-ZBLAN-Chalcogenide fibers
 & No & 1.5 - 10.5 $\mu$m & 50 mW & 3 MHz & 100 ps  \\
 \cite{Yan_OE_2021} & Butt Coupled ZBLAN-Chalcogenide fibers
 & Yes & 2 - 6.5 $\mu$m & 1.13 W & 600 kHz & 100 ps\\
 \cite{Petersen-OL-2019} & ZBLAN-Ge$_{10}$As$_{22}$Se$_{68}$ fiber taper
& Yes & 1.07 - 7.94 $\mu$m & 41 mW & Variable & 1 ns \\
\cite{Martinez_OL_2018} &  Silica-ZBLAN-As$_2$S$_3$-As$_2$Se$_3$ fibers
& Yes & 1.6 - 11 $\mu$m  & 139 mW & 800 kHz & 1.1 ns \\
 \cite{Petersen-SPIE-2016} & Silica-ZBLAN-As$_{38}$Se$_{62}$ fibers
 & Yes & 3.2 - 7.2 $\mu$m & 54.3 mW & 40 kHz & 3 ns \\
\cite{Gattas_OFT_2012} & HNLF-chalcogenide fibers&  Yes & 1.9 - 4.8 $\mu$m & 565 mW & 10 MHz & 40 ps \\

\end{tabular}
\end{table*}
On the other hand, among the most advanced cascaded fiber-based MIR SC systems, Venck et al. in 2020 demonstrated a MIR SC laser source with a bandwidth coverage between 2 to 10 $\mu$m in a cascaded silica-ZBLAN-As$_{38}$Se$_{62}$ fiber system as shown in Fig.\ref{CascadeSC} \cite{Venck_LPR_2020}. The experimental setup is shown in Fig.\ref{CascadeSC}(a), where a 25 m long ZBLAN fiber was directly pumped with a commercially available 460-ps pulsed fiber laser at 1.55 $\mu$m, without any fiber amplifier stage, that generated a SC spectra from 0.7-4.1 $\mu$m (see Fig. \ref{CascadeSC}(b)), whose short wavelength edge was then filtered down to 1.9 $\mu$m using a germanium anti-reflection coated long-pass filter. The filtered SC was then injected into a 9m long chalocogenide fiber that generated 2-10 $\mu$m SC (see Fig. \ref{CascadeSC}(c)) with an average output power of 16 mW \cite{Venck_LPR_2020}. This technique paves the way for low cost, practical, and robust broadband SC sources without the requirement of mid-IR pump sources or Thulium-doped fiber amplifiers. 

A list of most advanced and recently developed all fiber-based MIR SC systems are reported in Table \ref{table1}. In 2021, Woyessa et al. demonstrated a stable and portable MIR SC laser source covering the bandwidth from 1.46–10.46 $\mu$m with an average output power of 86.6 mW and a repetition rate of 3 MHz in a fiber cascade comprising of silica, ZBLAN, and chalcogenide fibers pumped by 0.5 ns pulsed Er/Yb-master oscillator power amplifier (MOPA) without any amplifier stages \cite{Woyessa_OL_2021}. Yan et al. demonstrated a watt-level (1.13 W) MIR SC spanning from 2-6.5 $\mu$m in a butt coupled ZBLAN and As$_2$S$_3$ cascaded fiber system pumped by a 1550 nm pulsed distributed feedback laser and few stages of thulium-doped fiber amplifier \cite{Yan_OE_2021}. In 2019, C. R. Petersen et al. reported 1.07-7.94 $\mu$m MIR SC with an average output power of 41 mW in an all-fiber cascade comprising of ZBLAN PCF and  Ge$_{10}$As$_{22}$Se$_{68}$ PCF taper. A variable repetition rate and 1 ns MOPA pumped ZBLAN PCF generated a SC from 1-4.2 $\mu$m, which was then injected into the chalcogenide taper that generated the long-wavelength edge of the SC up to 7.94 $\mu$m \cite{Petersen-OL-2019}. In 2018, Martinez et al. reported a MIR SC covering the bandwidth from 2 to 11 $\mu$m with 139 mW average output power in a cascaded fiber system using concatenated step-index ZBLAN, As$_2$S$_3$, and As$_2$Se$_3$ fibers. The fiber cascade was pumped at 1.553 $\mu$m by a MOPA system and three thulium-doped fiber amplifier stages \cite{Martinez_OL_2018}. In 2014, Kubat et al. theoretically reported SC generation spanning from 0.9 to 9 $\mu$m with an average output power of 15 mW above 6 $\mu$m in a ZBLAN-As$_2$Se$_3$ fiber cascade pumped by a thulium laser at 2 $\mu$m. The thulium laser pumped ZBLAN step-index fiber generated a SC from 0.9 to 4.1 $\mu$m, which is then injected into a 5 $\mu$m As$_2$Se$_3$ microstructured PCF that generated the long-wavelength edge \cite{Kubat_OE_2014}. C. R. Petersen et al. demonstrated a MIR cascaded SC generation beyond 7 $\mu$m in a silica-fluoride-chalcogenide fiber cascade pumped by a 1.55 $\mu$m seed laser and a thulium-doped fiber amplifier \cite{Petersen-SPIE-2016}. They pumped a commercially available Ge$_{10}$As$_{22}$Se$_{68}$-glass PCF with 135 mW of the pump continuum spanning from 3.5–4.4~$\mu$m and obtained a SC spectrum up to 7.2~$\mu$m with a total output power of 54.5~mW \cite{Petersen-SPIE-2016}. MIR SC spanning from 1.9- 4.8~$\mu$m with a record output power of 565~ mW was demonstrated by pumping a fiber cascade comprising of step-index silica and As$_2$S$_3$ fibers with a multistage MOPA system by Gattass et al. in 2012 \cite{Gattas_OFT_2012}. Commercially available MIR SC sources with an extended bandwidth up to 10~$\mu$m and tens of mW output power are now available. 

\subsection{Crystalline-core fibers}
In recent times, a new class of semiconductor fiber with crystalline core materials has emerged for nonlinear applications \cite{Peacock-SST-2016,Lagonigro-APL-2010}. These fibers are still clad in silica and retain some of the advantages of conventional glass fibers as they are robust, easy to handle, and can be post-processed using standard techniques. Although fibers with various core materials suitable for MIR SC generation have been fabricated, including Ge, SiGe, ZnSe, InSb, as of to date, the only semiconductor fibers with dimensions and transmission losses suitable for the observation of SC are those with a Si core \cite{Peacock-SST-2016,Lagonigro-APL-2010}. Compared to the chalcogenide-glass fibers, the Si core fibers offer a high nonlinear refractive index (~1.2 $\times$ 10$^{-17}$ m$^2$/W at wavelengths ~2~$\mu$m \cite{Bristow-APL-2007,Ren-OME-2019} and, thanks to the high core-cladding index contrast, a tight mode confinement \cite{Hsieh-OE-2007}, which is useful both for further increasing the nonlinear efficiency and for tailoring the dispersion. Coupled with their low losses in the MIR, these properties indicate the potential for Si core fibers to be used in SC generation in this region. This was confirmed by Ren et al. \cite{Ren-LSA-2019} when they demonstrated a high brightness and coherent MIR SC from 1.6-5.3 $\mu$m in a Si core fiber that had been asymmetrically tapered to optimise the nonlinear processing in the waist (2.8 $\mu$m diameter), whilst at the same time minimising the interaction of the long-wavelength light with the silica cladding at the output. The fiber was pumped at 3 $\mu$m with 100 fs pulses delivered from an OPO system \cite{Ren-LSA-2019}. 



\section{Ultraviolet range}
\label{UV}
A major challenge in SC generation is to overcome the shortcomings of existing fiber-based SC sources not only in the MIR but also in the UV wavelength range below 400 nm \cite{Travers_JO_2010,Sylvestre_OL_2012}. There is a particular need for broadband sources of UV light in applications such as multi-photon fluorescence microscopy for simultaneous coherent excitation of fluorescent proteins \cite{Poudel-JOSAB-2019}. However, UV generation in conventional silica-core fibers is extremely difficult because of many factors such as the strong material absorption, the photo-darkening effect, and the large normal dispersion. Tapering standard silica PCF has allowed for SC extension below 400 nm, yet limited to 320 nm \cite{Kudlinski_OE_2006,Travers_JO_2010}.

\subsection{Gas-filled hollow-core fibers}
The use of gas-filled hollow-core fibers for SC generation, especially towards the UV, has seen significant development over the last decade. There are multiple advantages to this approach~\cite{travers_ultrafast_2011,Russell_NP_2013,markosHybridPhotoniccrystalFiber2017}. Firstly, gases can be transparent from the vacuum ultraviolet (VUV) up to and across the infrared spectral range. The VUV guidance in particular extends far beyond what solid-core fibers can cover. Of course, the actual fiber structure must guide in these regions too. By using anti-resonant ``revolver'' style fibers, and similar, guidance in the VUV has been demonstrated~\cite{Belli_Optica_2015,ermolov_supercontinuum_2015,Winter:19,couchUltrafastMHzVacuumultraviolet2020a}. These guidance windows can extend significantly beyond the transmission region of the glass fiber structure itself, because in the anti-resonant regions the light is almost entirely contained in the hollow core. Secondly, gases can tolerate much higher intensities than glass, and ionized gas recombines and recovers, so that lasting damage is avoided. Therefore SC generation at much higher intensity is possible, opening the possibility of ultrafast SC applications which require high energy in a single shot. Thirdly, different gases have nonlinear properties which are different to solid-core fibers and tuneable. For example, noble gases do not exhibit Raman scattering, but do have ionisation and plasma related nonlinearity, leading to novel soliton dynamics. Alternatively, molecular gases exhibit Raman scattering, but with much narrower linewidth and larger frequency shifts compared to glass. Finally, and perhaps most importantly, the dispersion landscape and nonlinear coefficient can be tuned by simply changing the filling gas pressure or gas mixture. This provides a great degree of freedom for tuning the SC dynamics.

A wide range of nonlinear dynamics has been observed in gas-filled hollow-core fibers, reviewed previously~\cite{travers_ultrafast_2011,Russell_NP_2013,markosHybridPhotoniccrystalFiber2017}. Here we briefly summarize the main SC generation results. Many initial results targeted DW emission in the ultraviolet, but can be considered as SC generation by soliton fission~\cite{joly_bright_2011,Mak_OE_2013,Belli_Optica_2015,ermolov_supercontinuum_2015,Cassataro_OE_2017,Yu_OE_2018,Adamu_SCR_2019,Adamu_SCR_2020}, this includes important results on VUV SC generation~\cite{Belli_Optica_2015,ermolov_supercontinuum_2015}, with extension down to 113~nm achieved.  The VUV results made use of hollow-core PCF with core diameters around 30~$\mu$m, filled with either hydrogen or helium, pumped with $\sim$~30~fs pulses at 800~nm. With much longer pump pulses (500~fs) modulation instability based SC has also been demonstrated~\cite{tani_phz-wide_2013} in the visible and infrared region. While these have not yet been extended to the UV, numerical simulations suggest this should be feasible~\cite{travers_ultrafast_2011}. Recent novelties include utilizing the resonances in anti-resonant fibers for enhanced SC generation~\cite{sollapurResonanceenhancedMultioctaveSupercontinuum2017}, using tapered hollow-core fibers to enhance the ultraviolet generation~\cite{sureshDeepUVenhancedSupercontinuumGenerated2021}, and the conversion of a Raman frequency comb into a SC which also extended across the deep ultraviolet~\cite{gaoRamanFrequencyCombs2020}. 

There are also multiple disadvantages to using gas-filled hollow fibers. The fibers are incompatible with standard fiber technology, they require gas cells and gas and vacuum equipment to operate, they generally require higher power (and hence more expensive) lasers to pump them, and they are not widely available commercially.

\subsection{Novel glass fibers}
Although these results show great promise, compatibility with the ubiquitous silica platform remains a problem, and there is thus current interest in generating UV-light using modified UV-resistant glasses. Significant progress has been made to fabricate low-loss solarization-resistant UV-grade silica PCFs with high OH contents, enabling UV SC generation in the black-light region from 300 to 400 nm and beyond \cite{Perret_OSA_2020}. An alternative solution could be to use  ZBLAN-glass optical fibers. For instance, down to the UV (200 nm) to MIR SC has been generated by pumping into the cladding of a small-core ZBLAN PCF \cite{Jiang_NP_2015}. 

\section{Ultra-low noise and coherent SC sources}
\label{ANDI}


A detrimental feature of typical SC generation, i.e. pumping a fiber in the anomalous dispersion regime, is the low temporal coherence and highly stochastic spectral noise \cite{Wetzel_SC_2012}. Indeed, for injected pulses with soliton number $N > 16$, noise is stimulated through the nonlinear effects of modulation instability and Raman scattering \cite{Dudley_book_2010}. 

\subsection{All-normal dispersion (ANDi) fibers}
As an alternative approach, Heidt et al. investigated SC generation using high energy femtosecond pump pulses injected into ANDi PCFs, which exhibit low and flat normal dispersion over the entire bandwidth of the SC bandwidth \cite{Heidt2016, Heidt2010, Heidt_OE_2011, Hartung_OE_2011}. In this case, it was shown that only the nonlinear effects of SPM and OWB are responsible for the spectral broadening of the initial pulse (see Fig. \ref{ANDiSC}.(b)), which are both known to be highly coherent \cite{Finot_JOSAB_2008}.  Although the concept was introduced already over a decade ago, the strong application-driven demand for low-noise SC sources has resparked the scientific interest in the ANDi fiber design as it strongly suppresses the gain for noise-amplifying incoherent nonlinear dynamics \cite{Heidt_JOSAB_2017}. Here, we review the recent progress in the development of ultra-low noise SC sources and their remarkable impact on spectroscopy, imaging, and ultrafast photonics applications.

The superior coherence and noise properties of ANDi SC over conventional SC were experimentally verified, for example, by measurements of relative intensity noise (RIN), spectral coherence using unequal path Michelson interferometers, dispersive Fourier transformation, and RF beating with stabilized laser diodes \cite{Nishizawa2018, Klimczak2016, Tarnowski-SCR_2019, Rampur_APL_2021, Liu2015}. However, it is important to note that even the ANDi fiber design does not guarantee the generation of low-noise, coherent SC in every experimental configuration. The competition between coherent and quantum-noise amplifying incoherent nonlinear dynamics typically leads to a threshold pump pulse duration $T_{crit}$ or threshold input soliton number $N_{crit}$ above which the nature of the SC changes from coherent to incoherent. Indeed, as demonstrated in \cite{Heidt_JOSAB_2017}, for pump pulse durations longer than $\sim$ 1000 fs the contributions of SRS and FWM are not negligible anymore and start to degrade the SC coherence. Nevertheless, the ANDi fiber design exhibits about $10 \times$ higher $T_{crit}$ and $50 \times $ higher $N_{crit}$ than its conventional, anomalously pumped counterparts for octave-spanning bandwidths \cite{Heidt_JOSAB_2017}. Another limitation of ANDi SC generation is induced by the coherent coupling of the fiber polarization modes, which can lead to unpredictable fluctuations of the polarization state. This phenomenon is called PMI and is the vectorial aspect of the scalar MI. Experimental and numerical studies of PMI effects in ANDi SC generation were reported in weakly birefringent ANDi PCF \cite{Gonzalo_SCR_2018}, where the authors demonstrated that decoherence due to PMI starts to appear for pulse durations as short as 120 fs ($N_{crit} \simeq 30$). However, PMI effects can be effectively reduced or suppressed by using highly birefringent polarization-maintaining (PM) ANDi fiber designs \cite{Rampur_APL_2021,Liu2015}. Further instabilities related to incoherent polarization mode coupling and Raman amplification of noise were studied in \cite{Price-OE-2020, Price-JOSAB-2020, Gonzalo_JOSAB_2018}, with relevance mainly for pump pulse durations in the picosecond regime.


\begin{figure}[t!]
\centering
\includegraphics[width=8.8cm]{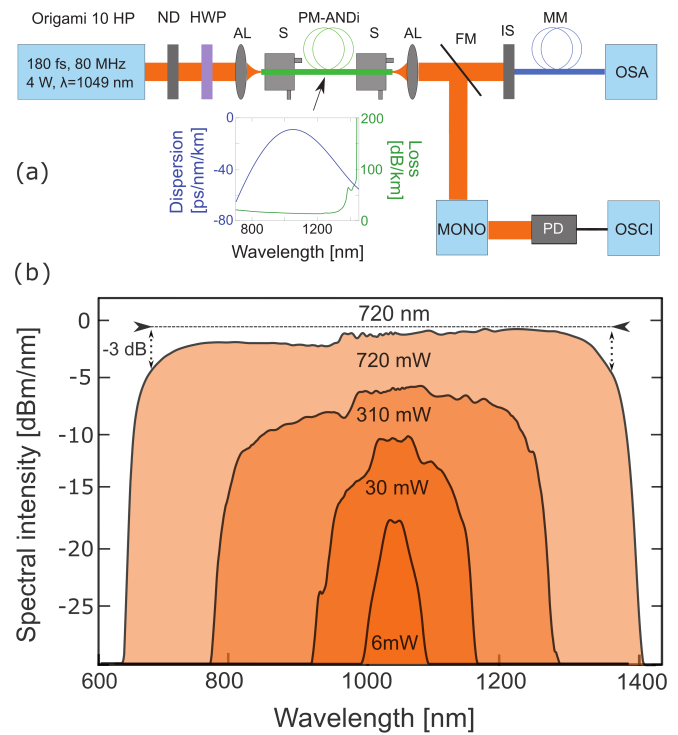}
\caption{(a) Experimental setup for ultra-low noise SC generation in a PM-ANDi PCF fiber, including ytterbium femtosecond mode-locked laser (ORIGAMI 10 HP), variable neutral density filter
(ND), half-wave plate (HWP), aspheric lenses (AL), 3D translation stages (S), 2 m of PM-ANDi PCF, flip-mirror (FM), integrating
sphere (IS), 2 m of multimode pick-up fiber (MM), optical spectrum analyzer (OSA), monochromator (MONO), photodiode (PD),
and oscilloscope (OSCI). The inset shows the fiber dispersion and its attenuation spectrum.
(b) Experimental SC spectra at the output of the PM-ANDi PCF for average output power of 6 mW to 720 mW with an input beam polarized along the fiber fast axis. Taken from Ref.~\cite{Genier-OL-2021}.}
\label{HP_SC_fast}
\end{figure}

In addition to these limitations induced by different nonlinear effects, the impact of technical noise on ANDi SC generation, such as RIN of the pump pulses at levels higher than the shot-noise limit, was the subject of several recent numerical and experimental studies \cite{Genier_JOSAB_2019, Rao-OL-2019, Heidt-OSAC-2020, Eslami2020}, with seemingly contradictory results. While it was demonstrated that a pump laser RIN as low as 0.6\% drastically degrades the coherence even for pump pulses as short as 100 fs \cite{Genier_JOSAB_2019}, the RIN of the SC in the central part of the spectrum can actually be lower than the RIN of the pump laser itself \cite{Genier_JOSAB_2019, Rao-OL-2019}. This result was unexpected but clearly explained by \cite{Heidt-OSAC-2020}, where the authors determined that the amplitude noise is converted into timing jitter, which affects only the coherence and not the RIN of the SC. However, this timing jitter remains in the order of only 100 attoseconds for practical pump laser parameters, which is up to two orders of magnitude lower than the jitter introduced by conventional soliton-based SC \cite{Hua2020, Rothhardt2012}, and therefore has little practical relevance for most applications. In fact, these studies have revealed the remarkable resilience of ANDi SC generation dynamics to technical pump laser noise under realistic pumping conditions.
\begin{table*}[h!]
\centering
\caption{\label{Table2} Summary of coherent SC generation in ANDi fibers and their noise (RIN) levels. $\lambda_{0}$ is the pump wavelength, $P_{0}$ the peak power, $T_{0}$ the pulse duration, $\Delta\lambda$ the SC bandwidth. PM - polarization maintaining. ESA - electric spectrum analyzer}
 \begin{tabular}{|c|c|c|c|c|c|c|c|c|c|}
 \hline
     \multicolumn{6}{|c|}{\textbf{SC properties}} & \multicolumn{4}{c|}{ \textbf{Noise properties}} \\ \hline
     Ref. & $\lambda_{0}$ & $P_{0}$ & $T_{0}$ & $\Delta\lambda$ & PM & Method & RIN bandwidth & Filter bandwidth & RIN  \\ \hline
         \cite{Heidt_OE_2011} & 1050 nm & 137 kW & 50 fs & 905 nm, -20 dB &  No & \multicolumn{4}{l|}{ \hspace{3.5cm} No Data} \\ \hline
    \cite{Heidt_OE_2011} & 790 nm & 220 kW & 50 fs & 880 nm, -20 dB  & No & \multicolumn{4}{l|}{ \hspace{3.5cm} No Data} \\ \hline
    
     \cite{Gonzalo_SCR_2018} & 1054 nm & 44 kW & 170 fs & 620 nm, -10 dB  & No  & Osci. & 600 nm & 10 nm & 33 \%\\ \hline
     \cite{Rao-OL-2019} & 1550 nm & 9 kW & 120 fs & 480 nm, -30 dB  & No & Osci. & 200 nm & 12 nm & 1.2 \% \\ \hline
     \cite{Genier-OL-2021} & 1049 nm & 48 kW & 180 fs & 720 nm, -3 dB  & 17 dB & Osci. & 400 nm & 3 nm & 0.54 \%   \\ \hline
    
    \cite{Resan-OE-2016} & 1040 nm & 120 kW & 230 fs & 600nm, -10 dB  & No  & ESA & 450 nm & 100 nm & 2 \%\\ \hline
        \cite{Chow-EL-2006} & 1562 nm & 3.3 kW & 170 fs & 185 nm, -30 dB  & No & \multicolumn{4}{l|}{ \hspace{3.5cm} No Data} \\ \hline
            \cite{Tarnowski_OE_2017} & 1800 nm & 400 kW & 70 fs & 1500 nm, -20 dB & 10 dB & \multicolumn{4}{l|}{ \hspace{3.5cm} No Data}\\ \hline
  \cite{Rampur-OE-2019} & 1560 nm & 22 kW & 80 fs & 1050 nm, -10 dB  & 10 dB & ESA & 1050 nm & no filter & 0.045 \% \\ \hline
\cite{Kasztelanic-OQE-2021} & 1030 nm & 62.5 kW & 400 fs & 500 nm, -60 dB & No & \multicolumn{4}{l|}{ \hspace{3.5cm} No Data} \\ \hline
\cite{Yuan2020} & 5000 nm & 160 MW & 150 fs & 11.2 $\mu$m, -30 dB  & No & \multicolumn{4}{l|}{ \hspace{3.5cm} No Data} \\ \hline
    \end{tabular}
\label{ANDi_recap}
\end{table*}

Fig. \ref{HP_SC_fast} shows recent measurements of SC generation in highly birefringent silica ANDi PCF, pumped by 180 fs pulses at 1049 nm \cite{Genier-OL-2021}. At high pump power, the combined action of SPM and OWB results in a spectrum with superb flatness covering the 670 - 1390 nm range. The SC exhibits high spectral power density up to 0.4 mW/nm, polarization extinction ratio of 17 dB, RIN as low as 0.54\%, and a simple temporal pulse shape with near-linear chirp.
Table \ref{ANDi_recap} further summarizes the characteristics of several ANDi SC sources reported in recent literature. For the realization of the different fiber designs, a wide range of instruments from the tool box of dispersion engineering in specialty optical fibers has been used, including air-hole microstructures \cite{Heidt_OE_2011, Gonzalo_SCR_2018, Genier-OL-2021, Rao-OL-2019, Resan-OE-2016, Chow-EL-2006}, microstructures combined with a highly germanium-doped core \cite{Tarnowski_OE_2017}, the integration of different soft glasses into all-solid PCF designs \cite{Rampur-OE-2019}, and selective infiltration of PCF holes with liquids \cite{Kasztelanic-OQE-2021}. Enabled by the high nonlinearity and transparency of chalcogenide glasses, the spectral coverage of ANDi SC sources has been extended far into the mid-IR to wavelengths up to 14 µm, e.g. using step-index, tapered or all-solid PCF designs \cite{Zhang2019, Yuan2020, Jiao2019, Nagasaka2017}. Where available, details of experimental RIN measurements are also listed, such as total bandwidth over which RIN was investigated, measurement method and spectral filter bandwidth \cite{Gonzalo_SCR_2018,Rao-OL-2019, Resan-OE-2016,Rampur-OE-2019, Genier-OL-2021}. In Refs. \cite{Gonzalo_SCR_2018} and \cite{Rao-OL-2019}, the authors used a setup composed of bandpass filters (10/12 nm), a fast photodiode and an oscilloscope to measure the spectrally-resolved SC RIN. In \cite{Gonzalo_SCR_2018},  a high average RIN of 33 \% was reported, which was attributed to SC coherence degradation by PMI in the weakly birefringent ANDi fiber.  In \cite{Rao-OL-2019}, a low average RIN of 1.2 \% was reported despite the use of a weakly birefringent fiber, owing to a low peak power of the pump pulses. Finally, in \cite{Genier-OL-2021}, the authors used highly birefringent, polarization-maintaining ANDi fiber to suppress PMI. They reported a flat spectrally resolved RIN level down to 0.54 \% over 400 nm using a monochromator with 3 nm filter bandwidth. Based on the summary in Table \ref{ANDi_recap}, this was the lowest average RIN value measured using the oscilloscope measurement method. While this is still roughly an order of magnitude larger than the result reported for an ANDi SC used to coherently seed a Tm:fiber amplifier \cite{Rampur-OE-2019}, it is important to note that these two RIN measurements cannot be directly compared since the latter was not spectrally resolved but measured over the entire bandwidth of the SC.



%
\subsection{Selected application examples of ANDi SC}
For over a decade, commercially available SC sources have made a tremendous impact in advanced spectroscopy and imaging techniques, such as multiphoton microscopy, stimulated emission depletion (STED) microscopy, OCT, and optical resolution photo-acoustic microscopy. Pumped by high repetition-rate pico- or nanosecond pulsed lasers and equipped with the high beam quality of optical fibers, the brightness of these table-top fiber-based SC sources even surpasses synchrotron beamlines \cite{Petersen_IPT_2018}. However, due to the stochastic nature of the nonlinear processes involved in spectral broadening using long pump pulses injected into the anomalous dispersion region of a nonlinear fiber, these SC sources provide spatially but not temporally coherent light and exhibit very large pulse-to-pulse fluctuations of spectral amplitude and phase \cite{Dudley_book_2010}. The fast-paced advancement of spectroscopic detection and imaging techniques has made this SC noise the predominating factor limiting acquisition speed, sensitivity, or resolution in many applications \cite{Jensen2019}. The adaptation of low-noise ultrafast ANDi SC sources therefore creates novel opportunities for applications in which the spectral uniformity, the temporal profile or the stability of the continuum is of importance and that have hence struggled to incorporate the noise-sensitive and complex conventional SC sources \cite{Heidt2016}. \\
\indent Hyperspectral SRS microscopy, which is used for label-free, chemical-specific biomedical and mineralogical imaging, is one example of such an application that would benefit from a broadband light source, but the high sensitivity to source noise has so far excluded the use of most nonlinear spectral broadening schemes. As recently demonstrated by Abdolghader et al. \cite{Abdolghader_OE_2020}, ANDi SC exhibit very low incremental source noise even when pumped with relatively long 220 fs pulses, and thus represent a viable and easily implementable low-noise source for broadband SRS imaging, greatly enhancing imaging speed in comparison to the alternative of using tunable narrowband lasers. ANDi SC sources were also the key-enabling technology behind the recent demonstration of spectrally resolved scanning near field optical microscopy (SNOM), where commercial SC sources are too noisy to be useful \cite{Kaltenecker2021}. The new technology enables the investigation of light-matter interaction with spatial resolution on the nanometer scale over a broad spectral bandwidth, relevant in particular to high-speed communications or quantum information processing. The low source noise was also the motivation of implementing ANDi SC in spectral-domain OCT for high axial resolution 3D-imaging, where it led to a paradigm shift as image quality is no longer limited by the SC noise but by detection shot noise. When compared to state-of-the-art commercial systems, ANDi-SC-based OCT imaging significantly enhances contrast, sensitivity, penetration, and speed, thus paving the way for improved medical diagnosis, e.g. in the early stage detection of skin cancer and other skin diseases \cite{Rao2021}. 

The full power of ANDi SC sources is harnessed by applications that require simultaneously broad spectral bandwidth, short pulses, and high coherence, such as single-beam coherent anti-Stokes Raman scattering (CARS) and other multi-photon microscopy techniques. In contrast to conventional SC, the underlying physics of ANDi SC generation preserve a single ultrafast pulse in the time domain, as shown in Fig.~\ref{ANDiSC}, enabling the delivery of fully compressed, broad-bandwidth, few-cycle pulses directly to the focus of highly dispersive microscope objectives. This is most efficiently accomplished by implementing adaptive pulse compression algorithms based on time-domain ptychography on a digital phase shaping device, such as a spatial light modulator, which has led to significant enhancement of signal-to-noise ratios, bandwidth and speed in CARS and multi-photon microscopy compared to alternative techniques \cite{Viljoen2020, Dwapanyin2020}. In addition, ANDi SC pulses with programmable spectral phase enable coherent control of the nonlinear response of the sample, reducing unwanted background and amplifying the otherwise weak resonant CARS signal \cite{Tu2014}. These advances have led to the development of an extremely versatile platform based on digitally programmable ANDi SC pulses, which combines multiple label-free nonlinear imaging and spectroscopy modalities into a single setup, offering the potential to translate this technology into routine clinical use for disease diagnosis \cite{Tu2016}. Even without the relatively expensive digital phase shaping device, broadband multimodal CARS and multi-photon imaging systems based on ANDi SC sources have recently been demonstrated \cite{Herdzik2020}. 

In ultrafast photonics, the unique temporal properties and high stability of ANDi SC pulses have been exploited for seeding ultra-broadband optical parametric chirped pulse amplification (OPCPA) systems, which have consequently facilitated the generation of coherent soft X-ray radiation and isolated attosecond pulses at high average powers and repetition rates \cite{Krebs2013, Rothhardt2017}. Light pulses with such extreme temporal and spectral properties are enabling new insights into the world of atoms, molecules, and their dynamics, e.g. via photoelectron spectroscopy or XUV microscopy \cite{Rothhardt2017}. On the other side of the light spectrum, the seeding of broadband ultrafast Thulium- and Holmium-doped fiber amplifiers operating at 2 $\mu$m wavelength with SC generated in highly birefringent ANDi fibers have resulted in an order of magnitude improvement of amplifier noise over comparable conventionally seeded implementations \cite{Rampur-OE-2019, Heidt2020}. This waveband is particularly important as stepping stone for the exploration of the molecular fingerprint region in the mid-infrared via nonlinear frequency conversion. Since recent experiments also convincingly demonstrated the advantages of ANDi SC over conventional SC in the construction of stabilized optical frequency combs for frequency metrology \cite{Nishizawa2018}, these studies have laid the foundations for the next generation of ultra-low noise frequency combs and ultrafast fiber amplifiers operating in the 2 $\mu$m spectral region and beyond in the MIR \cite{Rampur_APL_2021}. 


\section{Conclusions and Outlook}
\label{Conclusion}
The diverse results that have been summarized here clearly show that SC  generation in specialty optical fibers is a highly active field of research. We have specifically addressed recent challenges in extending the bandwidth of SC light sources towards both the UV and the MIR ranges in areas such as biomedical imaging and molecular spectroscopy. Extensive work over the last 15 years targeting the MIR range has resulted today in compact and reliable cascaded fluoride-chalcogenide fiber systems with watt-level output power over the 2 to 10~$\mu$m band. New research directions beyond 10~$\mu$m and up to 18~$\mu$m are being investigated using and Telluride-glass fibers \cite{Zhao_LPR_2017,Lemiere_RiP_2021}. 

While MIR SC fiber technology is becoming more mature, SC generation platforms in the UV and in the deep UV are still developing, and much effort is still needed before compact and reliable UV SC sources can be developed for direct applications. Of particular interest is the 315-400~nm black light (UV-A) band for applications in chemistry and medicine. UV-grade silica fibers are particularly interesting in this context as they can generate SC light in this UV-A band by pumping with UV picosecond lasers, but still with low output power and high losses. While SC generation in UV-grade fibers have seen compelling demonstrations, to date the most widely pursued form of UV SC generation is in gas-filled hollow core fibers, which dispenses with the silica absorption altogether. Such fibers can readily generate high-power broadband UV light down to 100 nm by pumping with femtosecond lasers.

Covering further the entire spectral range from the UV to the MIR and beyond into the Terahertz requires using scalable high-power few-cycle laser systems combined with difference-frequency generation (DFG) in nonlinear quadratic crystals. This has been recently and impressively achieved by Lesko et al. \cite{Lesko_NP_2021} and Elu et al. \cite{Elu_NP_2021}, who demonstrated record ultra-broadband SC comb generation from 350 nm to 22 $\mu$m and from 340 nm to 40 $\mu$m, respectively. 

We also reviewed the recent advances in ultra-flat, low-noise and coherent SC generation based on ANDi fibers. These results impressively illustrate the potential of ANDi SC sources for applications which so far were not able to use conventional, soliton-based fiber SC sources due to their noise or complex spectra and pulse shapes. 

Improved numerical modelling based on the GNLSE has been realized so far, which includes more fiber and pump laser parameters such as all types of noise sources (quantum noise, relative intensity noise, time jitter), pulse chirp, complete fiber absorption, and intermodal nonlinear interactions. Furthermore, machine learning analysis of SC generation appears also as a very promising topic towards improvement of SC bandwidth and optimization of both pump laser and fiber parameters \cite{Salmela_SC_2020}.

Finally, it is worth stressing here that the diverse numerical and experimental studies of SC generation in specialty fibers over the last 20 years have driven and enabled major parallel advances in developing nonlinear photonic integrated circuits (PICs) based on infrared glasses (TiO$_2$, AS$_2$S$_3$, Ta$_2$O$_5$) and other nonlinear materials (Si, Si$_3$N$_4$, SiGe, Ge) \cite{Lamont_OE_2008,Lau_OL_2014,Yu_LPR_2014,Singh_Optica_2015,Grassani_NC_2019}. Advances in micro-fabrication technology have led to the demonstration of SC generation in a number of PIC platforms, including SC covering the entire MIR first in a chalcogenide chip, and then in a silicon-germanium nanophotonic waveguide. The germanium-based platforms are particularly attractive due to their Complementary metal–oxide–semiconductor (CMOS) compatibility, their wide transmission window in the MIR and their large nonlinearities, enabling SC generation up to 13~$\mu$m \cite{Sinobad_Optica_2018,Montesinos_ACS_2020}. 

\noindent \textbf{Funding.}   Horizon 2020 Framework Programme (H2020) Marie Curie grant No. (722380 [SUPUVIR]); FP7 Information and Communication Technologies (ICT) (GALAHAD Project)  (7326); Agence Nationale de la Recherche (ANR) (ANR-15-IDEX-0003, ANR-17-EURE-0002, ANR-20-CE30-0004). Swiss National Science Foundation  (PCEFP2\_181222).

Disclosures. The authors declare no conflicts of interest.

\end{document}